\documentclass[
    ,final            
  ]
  {aipproc}

\layoutstyle{6x9}
\usepackage{amsmath,amssymb}
\usepackage{hyperref}
\usepackage[parfill]{parskip}
\usepackage{graphicx}

\begin{document}
\def\mod{\bf }    

\def\t{{t_0}}
\def\x{{\bf x}}
\def\y{{\bf y}}
\def\k{{\bf k}}
\def\A{{\it a}}
\def\ggamma{a}

\def\M{{\cal M}}
\def\a{\alpha}
\def\b{\beta}
\def\W{{W}}
\def\r{\rho}
\def\e{\eta}
\def\ex{e^\xi}
\def\emx{e^{m\xi}}
\def\x{{\bf x}}
\def\y{{\bf y}}
\def\k{{\bf k}}
\def\bra#1{\le\langle{#1}|}
\def\vrul{\rule[20pt]{0pt}{0pt}}
\def\ket#1{{#1}\ri\rangle}
\def\bra#1{\le\langle{#1},}
\def\bea{\begin{eqnarray}}
\def\eea{\end{eqnarray}}
\def\le{\left}
\def\ri{\right}
\def\l{\lambda}
\def\AdS#1{W^{(d+1)}_{\hbox{\scriptsize ${#1}$}}}
\def\Mink{W^{(d)}_{\hbox{\scriptsize \em M;}\sqrt{\lambda}}}
\def\RR{\mathbb R}
\def\C{\mathbb C}
\newtheorem{remark}{Remark}[section]
\def\bes{$$}
\def\ees{$$}
\def\beas{\begin{eqnarray*}}
\def\eeas{\end{eqnarray*}}
\def\PHI{{\widehat {\Phi}}}
\def\K{\varkappa}
\def\E{ E}
\def\P{ P}
\def\Sp{\hbox{Spec}(\square_\Y)}
\newtheorem{proposition}{Proposition}
\newcommand{\be}{\begin{equation}}
\newcommand{\ee}{\end{equation}}
\newcommand{\ba}{\begin{eqnarray}}
\newcommand{\ea}{\end{eqnarray}}
\newcommand{\pa}{\partial}
\renewcommand{\le}{\leq}
\renewcommand{\ge}{\geq}
\renewcommand{\Re}{\mathrm {Re} \,}
\renewcommand{\Im}{\mathrm {Im} \,}

\title{Quantum fields on curved spacetimes and a new look at the Unruh effect}


\author{Ugo Moschella}{
  address={Dipartimento di Fisica e Matematica. Universit\`a dell'Insubria, Como  \\ and INFN, sez.di Milano (Italy)}
}

\author{Richard Schaeffer}{
  address={Institut de Physique Th\'eorique, CEA-Saclay (France)}
}

\begin{abstract}

We describe a new viewpoint on canonical quantization of linear fields
on a general curved background that encompasses
and generalizes the standard treatment of canonical QFT given in textbooks.
Our method permits the construction of pure states and mixed stated with the same technique.
We apply our scheme to the study of Rindler QFT  and we present a new derivation
of the Unruh effect based on invariance arguments.

\end{abstract}

\maketitle


\section{First quantization: commutation relations and their representation}

Switching from classical to quantum mechanics  relies on a set of recipes that have
proven to successfully describe a large variety of physical phenomena, although the mysteries
of quantum mechanics and its interpretation persist to date.

In the simple case of a mechanical system having a finite number of degrees of
freedom,  the time-honored procedure of  ``canonical quantization''
essentially amounts to replacing the classical canonical variables $q_i, p_j$ and their Poisson brackets\footnote{The Poisson brackets of two functions of the canonical coordinates are defined as usual by
  \[ \{f,g\} = \sum \left(\frac{\partial f}{\partial q^{i}} \frac{\partial g}{\partial p_{i}}- \frac{\partial f}{\partial p_{i}} \frac{\partial g}{\partial q^{i}}\right)\]}  by noncommuting quantum variables $Q_i, P_j$ and their Canonical Commutation Relations (CCR's) \cite{heisenberg,dirac}:
\begin{eqnarray}
 \label{pbm} \{q_i,q_j\} = 0, \ \ \  \{p_i,p_j\} = 0, \ \ \  \{q_i,p_j\} = \delta_{ij} & & \ \ \ \  \makebox{(Poisson Brackets),} \\ && \cr [Q_i,Q_j] = 0, \ \ \ [P_i,P_j]= 0, \ \ \  [Q_i,P_j] = i \hbar \delta_{ij}  & & \ \ \ \  \makebox{(Commutators),} \label{ccrqp} \\ && \cr
  i,j = 1,\ldots, n & & \ \ \ \  n<\infty.
\end{eqnarray}
In a modern viewpoint the ``operators'' ${Q_i}$ and ${P_j}$ in Eq. (\ref{ccrqp}) can be regarded as elements of an abstract Heisenberg algebra rather than operators in a Hilbert space.
However, {\em under suitable technical assumptions},  the fundamental uniqueness theorem by Stone and von Neumann establishes that,  up to unitary equivalence,
there exists only one representation of the algebraic canonical commutation relations (\ref{ccrqp}) by
operators {in a Hilbert space $\cal H$:
\begin{eqnarray}
& & Q_i \to \hat Q_i \, , \ \ \  \ \  \ \  \   P_j \to \hat P_j\,: \label{rep}
\\ && \cr
&& [\hat Q_i,\hat Q_j] =0 ,\ \ \ [\hat P_i,\hat P_j] = 0 , \ \ \ [\hat Q_i,\hat P_j] = i \hbar \delta_{ij} {\bf 1} \label{rep1}
\end{eqnarray}
where the symbol } ${\bf 1}$ denotes the identity operator in the Hilbert space $\cal H$.
Therefore, the distinction between the abstract Heisenberg algebra and its representations
by means of operators in a Hilbert space is unnecessary for the study of finite-dimensional quantum mechanical systems.

As an elementary example, consider one pointlike particle in the physical  space  ${\Bbb R}^3$.
Following the probabilistic interpretation of quantum mechanics the natural Hilbert space to be considered is ${\cal H} = L^2({{\Bbb R}^3})$ and the commutation relations (\ref{ccrqp}) are  represented there by the following operators\footnote{We do not enter here into the unavoidable technical problems related to the unboundedness of operators; see e.g. \cite{reed} for an account of that important topic.}:
\begin{eqnarray}
\hat Q_i \psi(\x)  = x_i \psi(\x) , \ \ \ \ \hat P_j\psi(\x) = -ih \frac {\partial}{\partial x_j}\psi(\x), \ \ \ \ i,j= 1,2,3. \label{ccrqpreal}
\end{eqnarray}
The canonical commutation relations can be used to actually build the Hilbert space representation
starting from a ``fundamental'' state. To this end, one introduces the ladder operators \cite{dirac,messiah} as the following complex combinations of the elements of the Heisenberg algebra:
\begin{equation}\label{ccrau}
\A_i = \frac{Q_i + iP_i}{\sqrt 2}, \ \ \  \A_j^\dagger= \frac{Q_j - iP_j}{\sqrt 2}, \ \ \ i,j = 1,\ldots,n;
\end{equation}
the CCR algebra is rewritten as follows:
\begin{equation}\label{ccra}
[\A_i,\A_j] = 0, \ \ \ [\A_i^\dagger,\A_j^\dagger]= 0, \ \ \
[\A_i,\A^\dagger_j]= \hbar \delta_{ij}, \ \ \  i,j = 1,\ldots,n.
\end{equation}
A particularly relevant Hilbert space realization of the commutation relations (\ref{ccra})
\begin{equation}
 \A_i \to \hat \A_i, \ \ \ \A^\dagger_i \to \hat \A_i^\dagger ,
\end{equation}
\begin{equation}
[\hat \A_i,\hat \A_j] = 0, \ \ \ [\hat \A^\dagger_i,\hat \A^\dagger_j]= 0, \ \ \  [\hat \A_i,\hat \A^\dagger_j]= \hbar \delta_{ij} {\bf 1}, \ \ \  i,j = 1,\ldots,n, \label{craa}
\end{equation}
can be constructed starting
from a common eigenvector (the fundamental state) of all the annihilation operators $\hat \A_i $
with eigenvalue equal to zero:
\begin{equation}
\label{vacuum1}
\hat \A_i \left|0\right> = 0 \left|0\right>, \ \ \ i= 1,\ldots,n,
\end{equation}
with $\left<0 \right|\left. 0\right> = 1.$ An orthonormal basis of ${\cal H}$ is then obtained by repeated application of the creation operators to the fundamental state:
\begin{equation}
\label{nstates}\left|i_1,i_2,\ldots,i_n\right> \equiv ({{i_1!\, i_2!\ldots \, i_n!}})^{-\frac 12}(\hat \A^\dagger_1)^{i_1} (\hat \A^\dagger_2)^{i_2} \ldots (\hat \A^\dagger_n)^{i_n} \left|0\right>.
\end{equation}
A generic vector of the Hilbert space is then uniquely
identified is by a net of complex numbers $ c_{i_1 i_2 \ldots i_n} = \left<i_1,i_2,\ldots,i_n\right|\left.  \psi\right> $ such that
\begin{eqnarray}\label{not}
||\psi||^2 = \sum_{i_1,i_2,\ldots,i_N=0}^\infty |c_{i_1 i_2 \ldots i_n}|^2 < \infty
\end{eqnarray}
Matrix elements of the operators $\hat Q_i$, $\hat P_j$, $\hat \A_k$ and $\hat \A^\dagger_l$  can be easily computed in this representation by using the commutation relations (\ref{craa}) and the condition (\ref{vacuum1}). In particular the number operator
\begin{equation}\label{number}
N  = \sum_{i=1}^n \hat \A_i^\dagger \hat \A_i
\end{equation}
is diagonal. Every other Hilbert space
representations of the CCR satisfying the hypotheses of the Stone-Von Neumann theorem is unitarily equivalent to this one.

\section{Infinite Systems. Free fields in Minkowski space}

The situation drastically changes when considering systems
with infinitely many degrees of freedom: {the Stone-Von Neumann
theorem  is no longer applicable  and there exist
uncountably many {\it inequivalent } Hilbert space  representations of the canonical
commutation relations (\ref{ccrqp}) or (\ref{ccra}) when $n=\infty$ (see e.g. \cite{strocchi,haag}).
The distinction, then, } between a CCR algebra and a its Hilbert space representations
becomes crucial;  finding a quantum theory for an infinite system (such as a field)
involves therefore two distinct steps:
\begin{enumerate}
\item construction of an infinite dimensional algebra describing the degrees of freedom of the quantum system;
\item construction of a Hilbert space representation of that algebra.
\end{enumerate}
{ The final step would consist in choosing, among the infinitely many inequivalent representations,  a physically meaningful one.}
Unfortunately, a complete classification of the possible representations of
the canonical commutation relations does not  exist and is not foreseen in the near future.
This lack of knowledge is especially {relevant} in curved
backgrounds where, generally speaking, the selection of a representation
cannot be {guided by simple physical principles }as it is the case in flat space.
Indeed, while the CCR's have a purely kinematical content, the construction
and/or the choice of one specific representation {in a Hilbert space}
is always related to dynamics and different dynamical behaviors require
inequivalent representations of the CCR's (see e.g.
\cite{strocchi}; see however the discussion in \cite{halver}); in particular the choice of a specific representation
is related to many fundamental issues such as renormalizability, {
thermodynamical equilibrium and entropy,  symmetries, phase transitions, etc...}
The purpose of the following discussion is to
clarify some aspects the above non-uniqueness features by a detailed discussion of the
most elementary and well-known case: the Klein-Gordon field in Minkowski space.

\subsection{Canonical quantization of a Minkowski Klein-Gordon field}
\subsubsection{Classical field theory}
${M}_{d}$ denotes the $d$-dimensional Minkowski spacetime; an event $x \in M_d $
is parameterized by a set of
inertial coordinates $t=x^0,\ \x = (x^1,\ldots, x^{d-1})$.
The  nonvanishing covariant  components of the metric tensor are
$\eta_{0 0 } = - \eta_{ii} = 1$ in every inertial frame;
the Lorentz-invariant product of two events is given by
$x\cdot {x'}  = \eta_{\mu\nu} x^{\mu}{x'^{\nu}} = tt' - \x\cdot \x'$.

The dynamics of a real
massive Klein-Gordon scalar field $\phi(x)$
defined on ${M}_d$ is governed by the Lagrangian density
\begin{equation}\label{lagrangian}
{\cal L} = \frac 12 (\partial^\mu \phi) (\partial_\mu \phi) - \frac 12 m^2 \phi^2
\end{equation}
which implies the Klein-Gordon equation:
\begin{equation}
\Box \phi + m^2 \phi =  \partial^\mu \partial_\mu  \phi + m^2 \phi = \partial^2_t\phi - \Delta \phi + m^2 \phi= 0 . \label{kg}
\end{equation}
At a given time $\t$, the canonical variables encoding the degrees of freedom of the field  are the values of field itself
$\phi(\t,\x)$ and of the conjugated momenta
\begin{equation}\label{momentum}
\pi(\t,\x) = \frac{\partial {\cal L}}{\partial (\partial^t \phi)} = \partial_t \phi(\t,\x).
\end{equation}
{Here the spatial coordinates $\x$ are used as continuous labels for
the infinitely many degrees of freedom of the field in very much the same way as
the discrete indices $i$ of the finite dimensional case  label the degrees of freedom of the mechanical system (\ref{pbm}).
Consequently, the Poisson brackets of the canonical variables $\phi(\t,\x)$
and $\pi(\t,\x)$ can be formally assigned as in  (\ref{pbm}):
\begin{equation}\label{pbinf}
\{\phi(\t,\x), \phi (\t,\x')\} = 0, \ \ \{\pi(\t,\x), \pi (\t,\x')\}=0, \ \ \{\phi(\t,\x), \pi (\t,\x')\} =   \delta(\x-\x');
\end{equation}
the derivatives with respect to the variables $\phi$ and $\pi$ with continuous indices $\x$ or $\x'$ in the Poisson brackets are now functional derivatives. }
The interpretation of the above canonical structure is as follows.
Under suitable assumptions, the assignment of initial values at a given time $t=\t$
\begin{equation}\label{ini}
\phi(\t,\x) = f(\x), \ \ \pi(\t,\x) = g(\x)
\end{equation}
uniquely identifies a solution of the field equation (\ref{kg}).
This means that the phase space of the classical Klein-Gordon field theory can be identified
with the space of the possible initial condition at a given time, i.e.
a point of the phase space  is a pair  $[f,g]$ of  suitably smooth functions
$f,g:{\bf R}^{d-1} \to {\bf R}$ rapidly decreasing at infinity.
The linearity of the field equation (\ref{kg}) implies that the phase space also has a linear structure.
The symplectic structure of the phase space is encoded in the symplectic two-form
\begin{eqnarray}\label{symplectic}
\Omega ([f_1,g_1],[f_2,g_2]) &=& -\int_{{\bf R}^{d-1}} [f_1(\x) g_2(\x) - f_2(\x) g_1(\x)] d\x \cr
&=& -\int_{{\bf R}^{d-1}} [\phi_1(\t,\x) \pi_2(\t,\x) - \pi_1(\t,\x) \phi_2(\t,\x)] d\x,
\end{eqnarray}
that accounts for the formal Poisson brackets defined above; here the integral is overall the space manifold ${{\bf R}^{d-1}}$.
By using the Klein-Gordon equation (\ref{kg})
it is readily seen that the symplectic form does not depend on the time coordinate  $\t$:
\begin{eqnarray}\label{symplectic2}
\Omega (\phi_1,\phi_2) &=& -\int_{{\bf R}^{d-1}} [\phi_1(t,\x) \pi_2(t,\x) - \pi_1(t,\x) \phi_2(t,\x)] d\x
\end{eqnarray}
The  phase space can actually be  identified with
the linear space $\cal T$ of real solutions of the field equation (\ref{kg})
for initial condition belonging to a suitably chosen function space;
$\cal T$ is endowed with the symplectic form $\Omega (\phi_1,\phi_2)$.

\subsubsection{Quantization: CCR algebras}
Following the discussion of the finite-dimensional case,
in order to quantize the {classical theory with infinitely many degrees of freedom outlined in the previous section, we }introduce
a fixed time CCR algebra by imposing the following canonical commutation relations:
\begin{equation}\label{ccrinf}
[\phi(t,\x), \phi (t,\x')] = 0, \ \ \ [\pi(t,\x), \pi (t,\x')]=0, \ \ \ [\phi(t,\x), \pi (t,\x')] = i  \delta(\x-\x');
\end{equation}
we have used here the same letters $\phi$ and $\pi$ {as in the previous section} also
for the canonical quantum variables and set $\hbar = 1$.

A practical way to implement the CCR's (\ref{ccrinf})
goes as follows: the symplectic form $\Omega(\cdot,\cdot) $
can be extended to a pseudo-scalar
inner product in the linear space $\cal V$ of {\em complex} classical solutions of the
Klein-Gordon equation (\ref{kg}):
\begin{eqnarray}
(u_1,u_2) &=& i \int_{{\bf R}^{d-1}} {u}^*_1(t,\x)
\stackrel{\leftrightarrow}{{\partial_t} } u_2(t,\x)\  d\x\, =  \cr &=& i \int_{{\bf R}^{d-1}} {u}^*_1(t,\x)
\partial_t u_2(t,\x) - (\partial_t{u}^*_1(t,\x))
 u_2(t,\x)\  d\x\, . \label{pseudosc}
\end{eqnarray}
One then looks for set of solutions  $u_\k(t,\x)$  of
the Klein-Gordon equation  labeled by the spatial momenta
$\k$ dual to the ``spatial labels'' $\x$;
the set $\{  u_\k(t,\x)\}$ should be complete and orthonormal in the following sense:
\begin{equation}
(u_\k,u_{\k'}) = \delta(\k-\k'),\;\;\;\;( u^*_{\k},  u^*_{\k'}) = -\delta(\k-\k'),
\;\;\;\;(  u^*_\k, u_{\k'}) = 0.
\label{conditionsp}
\end{equation}
A real solution of the Klein-Gordon equation can therefore be written
as the following real superposition of the modes $u_\k(t,\x)$:
\begin{eqnarray}
\phi(t,\x) &=& \int \left[{u}^*_\k(t,\x)\,a^\dagger(\k) +  u_\k(t,\x)\, a(\k)  \right] d\k
\label{field0},
\end{eqnarray}
where the {elements of the CCR algebra $a_\k$ and $a_\k^\dagger$ are related to $\phi$ by}
\begin{eqnarray}
 a_\k =  (u_\k,\phi) , \ \ \  \  a^\dagger_\k  = {(u_\k,\phi)}^*.
\end{eqnarray}
The standard plane wave solution to these requirements is written as follows
\begin{equation}
u_\k(x)=u_\k(t,\x) = \frac{1}{\sqrt{2\omega(2\pi)^{d-1}}} \ \exp(-i\omega t +i\k\cdot \x),
\label{standpw}
\end{equation}
where
\begin{equation}
\label{omega}
\omega = \sqrt{\k^2+m^2}.
\end{equation}
These waves have ``positive frequency''
\begin{equation}\label{posfreq}
i \frac {\partial }{\partial t} \ u_\k(t,\x) = \omega \ u_\k (t,\x)
\end{equation}
i.e. they are eigenfunctions of the global timelike Killing
vector field $i \partial_t$ with positive eigenvalue $\omega$.
One easily checks {that the conditions (\ref{conditionsp})
are verified for the modes(\ref{standpw})}.
The field $\phi(x)$  and its canonically conjugated momentum
$\pi(x)$ admit therefore as following real mode expansions:
\begin{eqnarray}
\phi(x) &=& \frac{1}{{(2\pi)}^\frac{d-1}2}\int\left[ e^{i\omega t -i\k\cdot \x}\,\,a^\dagger(\k) +   e^{-i\omega t +i\k\cdot \x}\, a(\k)\right] \ \frac{d\k}{\sqrt{2\omega}}
\label{field1}\, ,
\\
\pi(x)& =&   \frac{i}{{(2\pi)}^\frac{d-1}2}\int\left[ e^{i\omega t -i\k\cdot \x}\,\,a^\dagger(\k) -
e^{-i\omega t +i\k\cdot \x}\, a(\k)\right] {\sqrt{\frac \omega 2}} \ {d\k}
\label{field2}.
\end{eqnarray}
Now, assume that the coefficients $a_\k$ and $a_\k^\dagger$ satisfy the
following  canonical commutation relations in  $\k$-space:
\begin{equation}\label{ccrinf2}
[a(\k),a({\k'})]= [a^\dagger(\k),
a^\dagger({\k'})] = 0,\ \ \ \ \ \ \ \ \ \ [a(\k),a^\dagger({\k'})] = \delta(\k-\k');
\end{equation}
{the fundamental  $\x$-space canonical commutation relations (\ref{ccrinf}) immediately follow from
(\ref{field1}) and (\ref{field2}). }

The equal-time CCR's  (\ref{ccrinf}) are equivalently expressed in a covariant
form by computing  the unequal-time commutator:
\begin{equation}\label{covcom}
[\phi(x),\phi(x')] = C(x,x') = \frac{1}{(2\pi)^{d-1}}
\int e^{-ik(x-x')} \epsilon(k^0)\delta(k^2-m^2)\,dk
\end{equation}
where
\begin{equation}\label{epsilon}
\epsilon(k^0) = \theta(k^0)-\theta(-k^0)
\end{equation}
and
\begin{equation}\label{theta}
\theta(k^0) = \left \{\begin{array}{ll} 1 & k^0>0\\ \frac 12 & k^0=0\\
0 & k^0<0 \end{array}\right.
\end{equation}
is the Heaviside's step function. The bi-distribution
$C(x,x')$ is a c-number; it is the unique bi-solution of the
Klein-Gordon equation with initial conditions (\ref{ccrinf}).

\subsubsection{Representing the quantum field in a Hilbert space I}
Up to this point we have constructed the quantum Klein-Gordon field
as a linear combination of the elements of the canonical algebra (\ref{ccrinf2}).
We have not yet represented the field and the canonical commutation relations
by operators  in a Hilbert space.

The first observation is that there exists a natural
representation $$\phi(x) \to \hat \phi(x)$$ (described at beginning of every book about QFT)
which is most obviously
related to the mode expansion (\ref{field1}) and which
can be constructed starting from the vacuum
vector annihilated by the (distributional) operators $\hat a(\k)$:
\begin{equation}
 \hat a(\k)\left |0_M \right>=0.\label{vacuum}
\end{equation}
The other vectors of the Hilbert space of this natural { -but by no means unique-} representation
 are constructed in perfect analogy with the finite dimensional case.
The vacuum vector $\left|0_M \right>$ is the (unique)
ground state for the energy operator and contains no particle;
the  $n$-particle states are constructed by repeated application of the creation operators
$\hat a^\dagger(\k)$ to the vacuum and  have positive energy in every Lorentz frame.
The two-point vacuum expectation value of the field ({\em in short: the two-point function
\footnote{For free fields a two-point function characterize completely a corresponding
Fock representation.
The Fock representation in turn can be constructed by applying the field algebra to
a certain reference cyclic vector often called the ``vacuum'' even when it is not void.
All these words (two-point function, representation, vacuum, state) are used interchangeably
 }}) has the
following expression:
\begin{eqnarray}\label{tp0}
&& W_0(x,x')= \langle 0_M | \hat \phi(x)\hat \phi(x')| 0_M\rangle =  \int {u}_\k(x){u}^*_\k(x')d\k = \cr &&=
\frac{1}{(2\pi)^{d-1}}\int  e^{-i\omega (t-t') +i\k\cdot (\x-\x')} \frac{d\k}{2\omega} = \frac{1}{(2\pi)^{d-1}}
\int e^{-ik(x-x')}
\theta(k^0)\delta(k^2-m^2)\,dk. \cr &&
\end{eqnarray}
Since we are dealing with a free field, the two-point function contains all the
information about the theory; this point
will be discussed later in detail.
One can easily verify that the covariant commutator coincides with  $C(x,y)$, see Eq.  (\ref{covcom}):
\[
 [\hat \phi(x),\hat \phi(x')] = \langle 0_M | [\hat \phi(x),\hat \phi(x')]| 0_M\rangle  =
\int [{u}_\k(x){u}^*_\k(x')-{u}^*_\k(x){u}_\k(x')] d\k = \]
\begin{equation}
= W_0(x,x') - W_0(x',x) =  C(x,x') = [\phi(x), \phi(x')]
\end{equation}

\subsubsection{Representing the quantum field in a Hilbert space II}

\noindent Any Hilbert space representation of the quantum field $\phi(x)$
must realize the equal-time CCR's (\ref{ccrinf}); equivalently,
the  covariant commutator of two field operators,
computed in the chosen representation,
must coincide with the distribution (\ref{covcom}).
Consider for instance the following change
of  basis in the canonical algebra (\ref{ccrinf2}):
\begin{equation}
\left\{\begin{array}{llrcl} \label{bogob}
b(\k) &=& \cosh \gamma(\k) \  a(\k) &-& \sinh \gamma(\k)\  a^\dagger(\k) \cr
b^\dagger(\k) &=& - \ \sinh \gamma(\k)\  a(\k) &+ & \cosh \gamma(\k)\  a^\dagger(\k)
\end{array}\right.
\end{equation}
It follows immediately that
\begin{equation}\label{ccrinf3}
[b(\k),b({\k'})]= [b^\dagger(\k),
b^\dagger({\k'})] = 0, \ \ \ \ \ \ \ \ \ \ [b(\k),b^\dagger({\k'})] = \delta(\k-\k').
\end{equation}
and the transformation is canonical.
Eq. (\ref{bogob}) defines a special instance of Bogoliubov
transformation. The (algebraic) field
\begin{eqnarray}
\phi(x) &=& \frac{1}{{(2\pi)}^\frac{d-1}2}\int\left[ e^{i\omega t -i\k\cdot \x}\,\,(\cosh \gamma(\k) \, b^\dagger(\k) + \sinh \gamma(\k)\, b(\k)) +  \right. \cr && + \left. e^{-i\omega t +i\k\cdot \x}\, (\cosh \gamma(\k) \, b(\k) + \sinh \gamma(\k)\, b^\dagger(\k))\right] \frac{d\k}{\sqrt{2\omega}}
\label{fieldb}\, \label{field4}
\end{eqnarray}
{by construction is the same field as the one given in (\ref{field1}); } it solves the Klein-Gordon equation and
provides an algebraic implementation of the covariant commutator (\ref{covcom}).
We can however use  the expression (\ref{field4}) to build another
representation
\footnote{We use the same symbol $\hat{} $   to indicate a generic operator in a Hilbert space without reference to any specific representation.}
 $\phi(x) \to \hat \phi(x)$;
starting  from the ``vacuum'' annihilated by the operators $\hat b(\k)$:
\begin{equation}\label{vacb}
{\hat b}(\k) \left|0_\gamma\right> = 0
\end{equation}
and proceeding as before. This representation is fully characterized by the two-point function
\vskip 16pt
$
 W_{(\gamma)}(x,x') =  \langle 0_{(\gamma)}| {\hat \phi}(x) {\hat \phi}(x')| 0_{(\gamma)}\rangle =
$
\begin{eqnarray}
\label{tpgamma}
&=& \frac{1}{(2\pi)^{d-1}}\int \left[ e^{-i\omega (t-t') +i\k\cdot (\x-\x')} \cosh^2 \gamma(\k) +  e^{i\omega (t-t') -i\k\cdot (\x-\x')} \sinh^2 \gamma(\k)\right]  \frac{d\k}{2\omega} + \cr
&+& \frac{1}{(2\pi)^{d-1}}\int \cos[\omega (t+t') -\k\cdot (\x+\x')]   \sinh (2\gamma(\k)) \frac{d\k}{2\omega}.
\end{eqnarray}
A straightforward computation shows that {the field $ \hat \phi(x)$ has the right covariant }commutator
\begin{eqnarray}
[{\hat \phi}(x),{\hat \phi} (x')]  = W_{(\gamma)}(x,x') -  W_{(\gamma)}(x',x) =  C(x,x')
\end{eqnarray}
and the quantization is canonical.
{When the Bogoliubov transformation (\ref{bogob})is not unitarily implementable
this construction gives rise to an inequivalent Hilbert
space representations  of the CCR's and of the field $\phi$.}
In particular, although the commutator is covariant, the Poincar\'e
symmetry is in general broken.
An example is provided by choosing $\gamma(\k)$
equal to a constant $\gamma$.
In this case the two-point function can be easily expressed in terms of $W_0(x,y)$
\begin{equation}
W_\gamma(x,x')= \cosh ^2\gamma\  W_0(x,x') + \sinh^2 \gamma \ W_0(x',x)
+ \frac{1}2\sinh 2\gamma\   [W_0(x,-x')+ W_0(-x,x')]. \label{nondiag}
\end{equation}
and the spacetime translation symmetry is manifestly broken.
We do not dwell here on the mathematical conditions that
discriminate between inequivalent representations and refer the reader to the literature.
The important point is that inequivalent representations describe
inequivalent physical situations (if any physical interpretation is available).

\subsubsection{Representing the quantum field in a Hilbert space III}

\noindent
Bogoliubov transformations are not the end of the story.
The fundamental $\x$-space CCR's (\ref{ccrinf})
that account for  the degrees of freedom of the quantum system
can be realized in another  way by making use of two independent
copies  of the momentum space CCR algebra (\ref{ccrinf2}).
Consider indeed a second independent canonical algebra
\begin{equation}\label{ccrinftilde}
[\tilde b(\k),\tilde b({\k'})]= [\tilde b^\dagger(\k),
\tilde b^\dagger({\k'})] = 0, \ \ \ \ \ \ \ \ \ \ [\tilde b(\k),\tilde b^\dagger({\k'})] = \delta(\k-\k'),
\end{equation}
and write
\begin{equation}
\left\{\begin{array}{llrcl} \label{hopf}
d(\k) &=& \cosh \gamma(\k) \  b(\k) &+& \sinh \gamma(\k)\  \tilde b^\dagger(\k) \cr
d^\dagger(\k) &=&  \ \sinh \gamma(\k)\  \tilde b(\k) &+ & \cosh \gamma(\k)\  b^\dagger(\k)
\end{array}\right. .
\end{equation}
It follows immediately that the momentum space CCR's  also hold for the newly introduced symbols:
$[d(\k),d^\dagger({\k'})] = \delta(\k-\k')$ etc..
Note however that the relations (\ref{hopf}) cannot be inverted.
The algebraic quantum field $\phi$
\begin{eqnarray}
\phi(x) &=& \frac{1}{{(2\pi)}^\frac{d-1}2}\int e^{i\omega t -i\k\cdot \x}\,\,[\cosh \gamma(\k) \ b^\dagger(\k) + \sinh \gamma(\k) \   \tilde b(\k) ] \ \frac{d\k}{\sqrt{2\omega}} +   \cr &+&  \frac{1}{{(2\pi)}^\frac{d-1}2} \int e^{-i\omega t +i\k\cdot \x}\, [\cosh \gamma(\k) \ b (\k) +  \sinh \gamma(\k) \  \tilde b^\dagger(\k)] \ \frac{d\k}{\sqrt{2\omega}}
\label{fieldhh}
\end{eqnarray}
is canonical and, as anticipated, the fundamental $\x$-space CCR's
(\ref{ccrinf}) have been implemented by using two independent copies of the $\k$-space CCR's (\ref{ccrinf2}).

The two-point function in the vacuum annihilated by the operators
$\hat b (\k)$  and $\widehat  {\tilde b} (\k)$ has the following expression:
\vskip10pt
$ \underline W_{(\gamma)}(x,x') = $
\begin{equation} = \frac{1}{(2\pi)^{d-1}}\int \left[ e^{-i\omega (t-t')
+i\k\cdot (\x-\x')} \cosh^2 \gamma(\k) +  e^{i\omega (t-t') -i\k\cdot (\x-\x')} \sinh^2 \gamma(\k)\right]  \frac{d\k}{2\omega}.
 \label{tpgammater}
\end{equation}

A  particularly important class of representations in the family of  states (\ref{tpgammater})
is obtained by choosing the following functions $\gamma(\k)$
\begin{equation}\label{beta}
\cosh \gamma(\k) = \sqrt{\frac{e^{\beta
\omega/2}}{2\sinh(\beta \omega)}};
\end{equation}
correspondingly
\begin{equation}
W_{\frac 1\beta}(x,x') = \frac{1}{(2\pi)^{d-1}}\int e^{-ik(x-x')} \left[\frac{1 }{1-e^{-\beta k^0}}\,\theta(k^0) + \frac{ 1} {e^{-\beta k^0}-1}\,\theta(-k^0)\right]\delta(k^2-m^2) dk.\label{wightther}
\end{equation}
These two-point functions are of fundamental
importance in quantum field theory as they
provide the Kubo-Martin-Schwinger (KMS) thermal representations
of the Klein-Gordon field at inverse temperature $\beta$
\cite{Birrell,kms,Bros:1992ey}.
Every inverse temperature $\beta$ describes a physically
distinct situation and the corresponding representations are inequivalent.
The ground state $W_{0}(x,x')$ is recovered in the limit $\beta \to \infty$.
One can appreciate here concretely the fact that inequivalent
representations of {\em the same canonical algebra}
have distinct physical meaning.

{The method of  doubling the CCR algebra as in Eq. (\ref{hopf})
is at the basis of the so-called thermofield theory \cite{umezawa}.
It may however be worthwhile to insist that
there is no doubling of the degrees of freedom
in representing one and the same field algebra (\ref{ccrinf}).
As we will see below, the doubling of the degrees of freedoms  is an
artifact of the momentum space representation
used to
implement the x-space CCR's.}

\section{Wightman-type approach}

For a free field, knowing the two-point function gives
all the information about the theory.
All the examples presented
in the previous section can be revisited by taking
the two-point function as the starting point.
In this section we will elaborate on this viewpoint in order
to prepare the discussion of the general curved case.

An acceptable two-point function of a Klein-Gordon field is a
positive-definite\footnote{The positive-definiteness hypothesis must however be relaxed when dealing with gauge QFT \cite{strocchi}}
distribution  $W \in {\cal S}' (M_d \times M_d)$
(or some other distribution space) that solves  the Klein-Gordon equation w.r.t. {both variables $x$ and  $x'$:}
\begin{equation}
\left(\square_x + m^2 \right)W(x,x')=\left(\square_{x'}
 + m^2 \right)W(x,x')
= 0\, . \label{kkgg}
\end{equation}
The condition of canonicity is written more conveniently
by using the covariant commutator:  $W(x,x')$ is required to be a solution of the functional equation
\begin{equation}\label{maincondition}
W(x,x') - W(x',x) = C(x,x').
\end{equation}
The Hilbert space of the theory ${\cal H}$ can be reconstructed
by using standard techniques \cite{swight,reed}.
The property of positive-definiteness of the two-point function is used to
defines a natural pre-Hilbert scalar product
in the space of test functions ${\cal S}(M_d)$ \cite{reed}:
\begin{equation}\label{scp}
(f,g) = \int_{M_d} f^*(x) W(x,x') g(x') dx \ dx'\ ,\ \ \ \ \ \ \ (f,f)\geq 0.
\end{equation}
The one particle Hilbert space  ${\cal H}^{(1)}$ is obtained by Hilbert completion of the quotient space ${\cal S}(M_d)/{\cal N}$
where ${\cal N}$ is the subspace of functions having  zero norm. The full Hilbert space of the model is the symmetric Fock space
${\cal H} = F_s({\cal H}^{(1)})$. As regards the  operator-valued distribution
\begin{equation}
\hat \phi : {\cal S}(M_d) \to Op({\cal H})
\end{equation}
representing the field, this is expressed  in terms of the ladder operators that
naturally act in the Fock space $\cal H$  (see Eq. ({\ref{azzone}}) below).
Any positive-definite two-point function solving (\ref{kkgg})
and (\ref{maincondition}) therefore
provides a canonical quantization of the Klein-Gordon field.
\subsubsection{Spacetime translation symmetry}
The additional hypothesis that the spacetime translations be an exact symmetry
amounts to requiring that the two-point function $W(x,x')$ depend only on the difference variable $\xi = x-x'$:
\begin{equation}\label{invast}
W(x,x') = W(x-x')= W(\xi).
\end{equation}
Eq. (\ref{kkgg}) is then most easily solved in Fourier space, where it becomes algebraic.
 The Fourier transform and antitransform of the distribution $W(\xi)$ are introduced as follows:
\begin{equation}
 \tilde W(k) = \int e^{ik\xi}W(\xi)  d\xi, \ \ \ \  W(\xi) = \frac{1}{(2\pi)^d}\int e^{-ik\xi} \tilde W(k)  dk;
\end{equation}
the Fourier representation of the Klein-Gordon equation (\ref{kkgg})
and of the functional equation (\ref{maincondition}) are respectively
\begin{eqnarray}
&& (k^2-m^2)\tilde W(k) = 0 \label{kgk}\, , \\
\label{maincondition2}
&& \tilde W(k) -\tilde W(-k) = \tilde C(k) = 2\pi\  \epsilon(k^0) \delta(k^2-m^2)\, \label{condk}.
\end{eqnarray}
However, translation invariance is not restrictive enough and there are yet infinitely many
possibly inequivalent solutions for these equations. They can be
parameterized by the choice of an arbitrary function $f(k)$ as follows:
\begin{eqnarray}\label{solutiongen}
\widetilde {W}_f(k) = 2\pi \, [f(k)+1]\theta(k^0)  \delta(k^2-m^2) + 2\pi \,f(-k)\theta(-k^0)  \delta(k^2-m^2);
\end{eqnarray}
$f(k)$ should be an acceptable multiplier for the distribution $\delta(k^2-m^2)$ so  that $\widetilde {W}_f(k)$
is a tempered distribution. Taking the Fourier anti-transform of $\tilde W_f(k)$
we obtain  the $x$-space representation of the two-point function:
\begin{equation}\label{solutiongenx}
W_{(f)}(x-x') = \frac{1}{(2\pi)^{d-1}} \int e^{-ik(x-x')} \left[(f(k)+1)\theta(k^0) + f(-k)\theta(-k^0)\right]  \delta(k^2-m^2)  dk
\end{equation}
The Bochner-Schwartz theorem tells that $W_f(x-x')$ is   positive-definite if and only if
the tempered distribution $\widetilde W_f(k)$ is a positive measure of polynomial growth.
This is in turn guaranteed by requiring that
\begin{equation}\label{nonn}
f(k)\geq 0
\end{equation}
and that the growth of $\widetilde W_f(k)$ is polynomial.
Eq. (\ref{solutiongen})
therefore provides a huge family of translation invariant canonical quantizations of the Klein-Gordon field.
Some physical criterium is needed  to discriminate among them: we will discuss a few examples.

\subsubsection{Ground state}

This representation is selected by imposing the physical
requirement of positivity of the spectrum
of the energy operator in every Lorentz frame.
Equivalently said, the joint spectrum of the energy-momentum vector operator $\P^\mu$
must be contained in the closed forward cone.
This condition implies that
the also the support of $\tilde W(k)$
must be contained in the closed forward
cone \cite{swight} and therefore the second term at the RHS of (\ref{solutiongen}) has to vanish i.e.
\begin{equation}\label{ll}
f(k) \equiv 0 \, .
\end{equation}
Positivity of the energy spectrum is therefore associated with the most obvious solution
of the functional equation (\ref{maincondition2}):
\begin{eqnarray}
\widetilde W_0(k) & = & 2\pi \theta(k^0) \delta(k^2-m^2). \label{wightk}
\end{eqnarray}
The $x$-space representation of two-point function (\ref{solutiongenx}) coincides precisely
with the expression already given in Eq. (\ref{tp0}).
The Fourier representation  (\ref{wightk}) also shows that the theory is invariant also under the restricted Lorentz group.

Finally, by plugging the integral representation (\ref{tp0}) into the scalar product (\ref{scp})
we obtain that the one-particle Hilbert space ${\cal H}^{(1)}$ is concretely realized in momentum space as follows:
\begin{equation}\label{hso}
{\cal H}^{(1)} = L^2\left( {\bf R}^{d-1}, \,  {d\k}/{\omega}\right).
\end{equation}
The relation between the two-point function $W_0$ and the positive frequency
solution of the Klein-Gordon equation can be further clarified by observing that for any test function $f\in {\cal S} (M_d)$, the convolution
\begin{equation}\label{conv}
    (W_0*f)(x) = \int W_0(x-x')f(x') dx'.
\end{equation}
is a smooth solution of the Klein-Gordon equation containing only positive frequencies.
The pseudo-scalar product introduced in (\ref{pseudosc})
is positive on the subspace generated by the positive frequency modes.
Hilbert completion of such subspace gives rise to the same one-particle space.
\subsubsection{Other {Poincar\'e-invariant} representation}
Dropping the requirement of positivity of the energy we immediately find a one parameter family
of other possible Lorentz invariant solutions characterized by the choice $f(k) \equiv \sinh^2 \gamma = $ constant:
\begin{eqnarray}
\widetilde {\underline W}_\gamma(k) & = & 2\pi [\cosh^2 \gamma \theta(k^0) + \sinh^2 \gamma \theta(-k^0)] \delta(k^2-m^2). \label{wight1k}
\end{eqnarray}
\begin{equation}
\underline{{W}}_\gamma(x,x')= \cosh ^2\gamma\, W_0(x,x') + \sinh^2 \gamma W_0(x',x)  \label{nondiag2}
\end{equation}
which is a canonical local and {Poincar\'e-invariant} solution of the Klein-Gordon equation.
Of course $W_\gamma$ does not satisfy the {spectral condition since states with negative } energy are now present.

\subsubsection{KMS representations}
Another split of the commutator is obtained
by replacing  {the Heaviside step function in (\ref{solutiongen}) by the Bose-Einstein factor}   as follows \cite{Bros:1992ey}:
\begin{equation}\widetilde W_\frac 1 \beta(k) =  \frac{2\pi}{1 - e^{-\beta k^0}}  \epsilon(k^0) \ \delta(k^2-m^2)\end{equation}
so that
\begin{eqnarray}
\widetilde W_\frac 1 \beta(-k) = \frac{2\pi}{1 - e^{\beta k^0}}  \epsilon(-k^0) \ \delta(k^2-m^2)= \frac{2\pi e^{-\beta k^0}}{1-e^{-\beta k^0}} \epsilon(k^0) \  \delta(k^2-m^2)
\end{eqnarray}
and the condition (\ref{maincondition2}) is manifestly satisfied.
Fourier antitransforming  $\tilde W_\beta(k)$ brings us back to the
two-point function given in Eq. (\ref{tpgammater}).
Furthermore, since $\tilde W_\beta(k)$ is a positive measure and therefore
$W_\beta(x-x')$ is positive definite.

At this point we  step back and observe
that the difference between the two-point functions  (\ref{tpgamma}) and (\ref{tpgammater})
is just the last term  at the RHS of Eq. (\ref{tpgamma}){
\begin{equation}
F(x,x') = \frac{1}{(2\pi)^{d-1}}\int \cos[\omega (t+s) -\k\cdot (\x+\x')]   \sinh (2\gamma(\k)) \frac{d\k}{2\omega}.
\end{equation}
which is absent in Eq. (\ref{tpgammater}). } Two remarks are in order here:

1) since both theories are canonical the contribution of $F(x,x')$ to the commutator  must be zero.
This is indeed true because $F(x,x')=F(x',x)$.

2) Adding any symmetric positive  function to an acceptable two-point function produces another acceptable two-point function
(provided positive-definiteness continue to hold.

For some purposes it may be useful to rewrite the two-point functions (\ref{tpgamma}) and (\ref{tpgammater}) as follows
 \begin{eqnarray}\label{qwerty}
 W_{(\gamma)} (x,x') &=& \int [u_{\k}(x) \ , \  u^*_{\k}(x)] \left[\begin{array}{ccc} \a_\k \a^*_\k& ,& \a_\k \b^*_\k \\
  \a^*_\k \b_\k & ,&  \b_\k \b^*_\k \end{array}\right]
  \left[\begin{array}{l} u^*_\k(x') \\
  u_\k(x') \end{array}\right] d\k\,
\cr
&=& \int [\a_\k u_{\k}(x) + \b_\k u^*_{\k}(x)] [\a^*_\k u^*_{\k}(x') + \b^*_\k u_{\k}(x')]
  d\k\, ,
\cr
&&
\\
\label{awerty}
 \underline W_{(\gamma)} (x,y) &=& \int [u_{\k}(x) \ , \  u^*_{\k}(x)] \left[\begin{array}{ccc} \a_\k \a^*_\k & ,& 0 \\
  0 & ,& \b_\k \b^*_\k  \end{array}\right]
  \left[\begin{array}{l} u^*_\k(x') \\
  u_\k(x') \end{array}\right] d\k\,
  \cr
&=& \int [\a_\k u_{\k}(x) \,\, \a^*_\k u^*_{\k}(x') + \b_\k u^*_{\k}(x)\, \, \b^*_\k u_{\k}(x')]
  d\k\,
  ,
 \end{eqnarray}
 with $\a_\k \a^*_\k- \b_\k \b^*_\k=1$.
Note the absence of the off diagonal terms in the second expression
$\underline W_{(\gamma)}$.

\section{Extended Canonical Formalism}
\label{exten}

\subsubsection{General Scheme}
We  now  summarize the findings and the hints of the previous section
 in a general framework suitable to
discuss the problem of quantizing a
field on a curved spacetime.
What follows is an  account of an extension of the
quantum canonical formalism that we have recently
introduced in \cite{{Moschella:2008ik}}.
We limit the discussion to the case of
a real Klein-Gordon field $\phi(x)$ on a
Lorentzian manifold ${\cal M}$.

A quantum field is a distributional map from a space
of test functions ${\cal T}({\cal M})$ into the elements
of a field algebra $\cal F$:
\begin{equation}
\phi: {\cal T}({\cal M}) \rightarrow {\cal F}.
\end{equation}
The algebraic structure is given by the commutation relations: the commutator has a purely algebraic or kinematical
content and, for free fields, is a multiple of the identity element of the algebra:
\begin{equation}[\phi(f), \phi(g)] = C(f,g)\,  = \int C(x,x')f(x) g(x') \, dx \, dx'\, ;
\label{comm}\end{equation}
 $C(x,x')$ is an {\em antisymmetric} bidistribution on the manifold ${\cal M}$, solving the Klein-Gordon equation in each variable and vanishing coherently with the notion of locality inherent
to ${\cal M}$, i.e.
$C(x,x') = 0$  for  any two events    $x,x' \in {\M}$ {spacelike separated}; $dx$ shortly denotes  the invariant volume form.

Quantizing means representing
the above commutation relations by an operator-valued distribution in a
Hilbert space $\cal H$ {\em i.e.}
\begin{equation}
\phi (f) \longrightarrow \hat \phi (f)\in Op({\cal H}) \ \ \
\makebox{such that} \ \ \ [\widehat \phi (f),\widehat \phi (g)] = C(f,g)\, {\mathbf 1}. \label{poi}
\end{equation}
For free fields the truncated $n$-point functions vanish and the theory is completed encoded in
the knowledge of a positive
semi-definite two-point function ${\W}(x,x')$, a distribution
whose interpretation is the two-point "vacuum" expectation value of the field:
\begin{equation}
{\W}(x,x') \equiv \langle \Omega, \hat \phi(x) \hat \phi (x') \Omega\rangle. \label{tp0d}
\end{equation}
As discussed in the previous sections, the following is the crucial property that ${\W}(x,x')$ has to satisfy to induce a Hilbert space representation of the commutation rules (\ref{comm}):
${\W}(x,x')$ must  realize a splitting of the commutator $C(x,x')$ by solving the following fundamental functional equation:
\begin{equation}
{\W}(x,x')- {\W}(x',x) = {C}(x,x').  \label{CR}
\end{equation}
Given a $\W$ satisfying (\ref{CR}), a Hilbert space representation
of the algebraic covariant CCR (\ref{comm}) can be constructed explicitly.
As in the flat case,
the one-particle space ${\cal H}^{(1)}$ is obtained by the standard Hilbert space completion
the space of test function ${\cal T}(\cal M$) w.r.t.
the  pre-Hilbert product provided by the two-point function:
\begin{equation}\label{scp2}
(f,g) = \int_{\cal M} f^*(x) W(x,x') g(x') dx \ dx'\ ,\ \ \ \ \ \ \ (f,f)\geq 0.
\end{equation}
The full Hilbert space of the theory is the  symmetric Fock
space ${\cal H}=  F_s({\cal H}^{(1)})$.
Each field operator $\widehat\phi(f)$ can be decomposed into
``creation'' and ``annihilation'' operators
defined by their action on the dense subset
of  $F_s({\cal H}^{(1)})$ of elements the form ${h} = ({h}_0, {h}_1,\ldots
{h}_n,\ldots,0, 0,0,\ldots)$:
\begin{equation}
\hat\phi(f)=\hat\phi^{+}(f)+\hat\phi^{-}(f) \label{azzone}
\end{equation}
\begin{eqnarray}
\left({\hat{\phi}}^{-}(f){h}\right)_{n}(x_1,\ldots,x_n)=
{\sqrt{n+1}}\int {\W}(x,x') f(x)
{h}_{n+1}(x',x_1,\ldots,x_n)
dx dx' ,
\\
\left(\hat\phi^{+}(f){h}\right)_n(x_1,\ldots,x_n)=\frac{1}{\sqrt{n}}
\sum_{k=1}^{n}f(x_k){h}_{n-1}(x_1,\ldots, \hat{x}_{k},\ldots,x_n).
\label{azzo}
\end{eqnarray}
Because of (\ref{CR}) these formulae imply the commutation relations (\ref{poi}).
Therefore, also in the curved case, quantizing a free theory
amounts to specify a two-point function solving (\ref{CR}).
Furthermore, all the  Green's functions can be constructed from the knowledge
of the two-point function (\ref{tp0d}) which therefore encodes
the dynamics of the free field and
the possibility of studying interacting models perturbatively.
\subsubsection{Canonical quantization: constructing a general two-point function}

The central hypothesis founding the theory of quantum fields on curved backgrounds is the globally
hyperbolicity\footnote{Note however that the hypothesis of global hyperbolicity is not satisfied by
physically interesting and important examples
like the anti-de Sitter spacetime, or in presence of a naked singularity. These
examples have to be treated by appropriate methods.
Our generalization of the canonical formalism may however work also in these cases. }
of the background manifold $\cal M$, which is assumed to admit a Cauchy surface $\Sigma$ (see e.g. \cite{wald}).
This implies that ${\cal M}$  is actually foliated by  spacelike Cauchy hypersurfaces,
\begin{equation}\label{ghyp}
{\cal M} = \bigcup_t \Sigma_t
\end{equation}
where $t$ is a global time coordinates.

Global hyperbolicity guarantees that the Cauchy problem for the Klein-Gordon equation is well-posed and that
the phase space  for the classical Klein-Gordon field theory can be identified
with the linear space of real solutions of the field equation.
There is a natural conserved
symplectic form defined on such space of real solutions, which, furthermore,
can be  extended  to an inner product in the space $\cal V$ of complex
solutions of the Klein-Gordon equation:
given  $u_1$ and $u_2 \in {\cal V}$ their inner product is defined as the integral of the conserved
current
\begin{equation}
j_\mu(x) = i \ {u^*_1(x)} \mathop{{\nabla}_\mu} u_2(x) -
i \left[ \mathop{{\nabla}_\mu}{u^*_1(x)} \right]  u_2(x)
\end{equation}
over a three-dimensional Cauchy surface:
\begin{equation} (u_1,u_2) = i\int_{\Sigma_t}  u_1^* \stackrel{\leftrightarrow}{\nabla}_\mu
u_2 \, d \sigma^\mu \, ;  \label{kgproduct}
\end{equation}
in this formula $d \sigma^\mu = n^\mu \sqrt{|h|} d^3x$ in local coordinates
on $\Sigma_t$, $h_{ij}$ is the  induced metric on $\Sigma_t$,
$n^\mu$ is a future-directed unit normal vector field on $\Sigma_t$.

The pseudo-scalar product (\ref{kgproduct}) generalizes (\ref{pseudosc})
to the curved case and is often called the Klein-Gordon product;
it is a conserved quantity i.e. it
does not depend on the choice of the Cauchy surface where the integral is performed.

A standard canonical quantization is then achieved \cite{wald} by finding a splitting of the above space of complex solutions
into the direct sum of a subspace where the inner product is positive
and of its complex conjugate.
The ``1-particle'' Hilbert space of the theory is finally obtained
by completing the positive subspace in the pre-Hilbert topology defined by the inner product restricted to it.
In favorable cases \cite{fulling}
one can realize the splitting by finding a family  $\{u_i(x)\}$
of  complex solutions complete and orthonormal
in the following sense
\begin{equation}
(u_i,u_j) = \delta_{ij},\;\;\;\;( u^*_i,  u^*_j) = -\delta_{ij};
\;\;\;\;(  u_i, u^*_j) = 0; \label{kgbase}
\end{equation}
the index $i$ generally refers to the spectrum of some elliptic operator appearing in the Klein-Gordon operator; for instance,
in the Minkowski case (\ref{conditionsp}) the index  $i=\k$ was labeling the eigenfunctions of the Laplace operator $\Delta$.

The standard procedure for canonical quantization can be then rephrased and summarized in the following two steps:
\begin{enumerate} \item
in the first step  the commutator is concretely represented as the following antisymmetric combination of the modes:
\begin{equation}
[\phi(x),\phi(x')] = C(x,x') = \sum_i [u_i(x) u_i^*(x') - u_i(x') u_i^*(x)]; \label{commu}
\end{equation}
the fact that $\{u_i(x)\}$ is a ``complete'' set of solutions of the Klein-Gordon equation
guarantees the CCR's;
\item
in the second step  the fundamental functional equation (\ref{CR})
is solved in the simplest way, coherently with the choice of orthonormal basis made in
(\ref{kgbase}) and used in (\ref{commu}):
\begin{equation}
\W(x,x') = \langle \Omega, \hat \phi(x) \hat \phi(x')  \Omega \rangle = \sum u_i(x) u_i^*(x'). \label{standard}
\end{equation}
Positive-definiteness of the two-point function (\ref{standard}) is manifest.
\end{enumerate}
The above description is equivalent to the usual
presentation of canonical quantization described in textbooks \cite{Birrell,wald}
and used in the vast majority of the enormous and still growing literature where quantum fields in curved backgrounds are relevant.
Indeed, the first step amounts to writing
the formal expansion of the field
\begin{equation}
\phi(x) = \sum[ u_i(x)a_i + u^*_i(x)a_i^\dagger]
\end{equation}
in terms of the elements of a space CCR algebra indexed by the same labels parametrizing the basis (\ref{kgbase}):
\begin{eqnarray}
&& [a_i,a^\dagger_j] = \delta_{ij}, \ \ \ \ \ \ [a_i,a_j]=0 , \ \ \ \ \ \ [a^\dagger_i,a^\dagger_j]= 0;
\end{eqnarray}
in the second step a Fock representation $\phi(x) \to \hat \phi(x)$ is specified by the conditions
\begin{equation}
\hat a_i |\Omega\rangle = 0,      \ \ \ \forall i.
\end{equation}
This representation coincides precisely the Fock representation (\ref{azzone})
constructed with the help of the two-point function (\ref{standard}).

A considerable amount of work has been devoted to the attempt
of formulating various alternative  prescriptions
to select, among the possible canonical quantizations constructed this way, those which
can have a meaningful physical interpretation. We refer the reader to the literature (see e.g. \cite{wald}).

There are however infinitely many other solutions of the functional equation (\ref{CR}) giving rise to other (possibly)
inequivalent canonical quantizations and we will now show how to construct many of them
by using the complete set (\ref{kgbase}). Following the  discussion of the flat case, we start once more by a rapid
review of the theory of Bogoliubov transformations.

A Bogoliubov transformation amounts to the construction of a second complete system $\{v_i(x)\}$  by
the specification of two complex operators (matrices) $\a_{ij}$ and $\b_{ij}$ such that
\begin{eqnarray}
v_i(x) &=& \a_{ij} u_j(x) + \b_{ij}  {u}^*_j(x) \label{bogdir}.
\end{eqnarray}
The transformation should of course  be invertible and the inverse given by
\begin{eqnarray}
 u_j(x) &=&  v_i(x)\, \a_{ij}^*  - {v}^*_i(x)\, \b_{ij}  \label{boginv}.
\end{eqnarray}
By composing  the direct and inverse transformations it follows  that
 $\a$ and $\b$ must satisfy the following conditions:
\begin{equation}
\a_{il} \a^*_{jl} - \b_{il} \b^*_{jl} = \delta_{ij}, \ \ \ \ \ \ \ \  \a_{il} \b_{jl} - \b_{il} \a_{jl} = 0,
\end{equation}
\begin{equation}
\a_{li}^* \a_{lj}  -\b_{li}  \b^*_{lj} = \delta_{ij}, \ \ \ \ \ \ \   \a^*_{li} \b_{lj} - \b_{li}
\a^*_{lj} = 0. \label{bb}
\end{equation}
The standard Fock quantization based on the system $\{v_i(x)\}$ is then encoded in the two-point
function
\begin{eqnarray}
\W_{\a,\b}(x,x') = \sum v_i(x) v_i^*(x') =  \sum\, [\a_{ij
} \, \a^*_{il} u_j(x)  u^*_l(x') + \b_{ij} \b^*_{il} \, u^*_j(x) \, u_l(x') \cr + \a_{ij
} \b^*_{il}\,  u_j(x)  u_l(x') + \b^*_{ij} \a_{il} \, u^*_j(x) \, u^*_l(x')]   \label{kgab}
\end{eqnarray}
interpreted as the two-point vacuum expectation value of the quantum field.
Positive definiteness of (\ref{kgab}) is evident.

When  $\b$ is a Hilbert-Schmidt operator this quantization turns out to be unitarily equivalent
to the Fock quantization (\ref{standard}). Otherwise (\ref{standard}) and (\ref{kgab}) give rise to inequivalent quantizations.
The commutator must however be independent of the choice of
$\a$ and $\b$; condition (\ref{bb}) precisely implies that this is true:
\begin{eqnarray}
\W_{\a,\b}(x,x')- \W_{\a,\b}(x',x) = \sum [v_i(x) v^*_i(x')-v_i(x')
v^*_i(x)] = \cr =\sum [u_i(x) u^*_i(x')-u_i(x')
u^*_i(x)] = C(x,x'). \label{cancomm} \end{eqnarray}
Eq. (\ref{kgab}) defines the most general two-point function,
solving the Klein-Gordon equation and functional equation
expressing the CCR (\ref{CR}), that can be constructed within the theory of
Bogoliubov transformations starting from the basis (\ref{kgbase}).

At this point in most textbooks the story about canonical quantization
comes to an end. There however exists a wealth of other representations of the field algebra, having important physical meaning,
that escape the above scheme.

We have shown  in \cite{Moschella:2008ik} that many new representations
(including the thermal equilibrium states) can be produced by a simple generalization
of the family of two-point functions displayed in (\ref{kgab}).
Consider indeed two hermitian matrices $A$ and $B$ and a
complex matrix $C$ and construct the general quadratic form
\begin{eqnarray}
Q(x,x') =  \sum [A_{ij} \, u_i(x) u^*_j(x') + B_{ij} \, u^*_i(x)
u_j(x') + C_{ij} u_i(x) u_j(x') + C^*_{ij} u^*_i(x)
u^*_j(x')].\end{eqnarray}
Now we impose that $Q(x,x')$ be a solution of the functional equation (\ref{CR})
as follows
(\ref{kgbase}):
\begin{equation}\label{cr2}
Q(x,x')-Q(x',x) = C(x,x') = \sum [u_i(x) u_i^*(x') - u_i(x') u_i^*(x)];
\end{equation}
we obtain   the operators $A, B$ and $C$ must satisfy the following conditions:
\begin{equation}
A_{ij} -
B_{ji} = \delta_{ij}, \ \ \ \ C_{ij} - C_{ji} = 0.
\end{equation}
The most general expression of a two-point function
solving the Klein-Gordon equation and the the functional equation (\ref{CR}) is therefore given by
\begin{eqnarray}
\W(x,x') &=& \sum  \left[ \delta_{ij} + B_{ji}\right] u_i(x) u^*_j(x') + \sum B_{ij} u^*_i(x) u_j(x') \cr
&+&  \Re \sum C_{ij}
[u_i(x) u_j(x') + u_i(x') u_j(x)] + S(x,x'). \label{general}
\end{eqnarray}
{Further restrictions are  imposed by the requirement of positive definiteness
for the form (\ref{general}). }
Only the first diagonal term at the RHS
contributes to the commutator. The other terms
altogether constitute the most general combination of the modes
(\ref{kgbase}) so that the
total contribution to the commutator vanish. The additional term
$S(x,x')$ which  appear at RHS is a bisolution  of the Klein-Gordon equation
that is not ``square-integrable'' (even in a generalized sense).
It is of classical nature
and symmetric in the exchange of $x$ and $x'$.
Quantum constraints do not generally forbid the existence of
such a contribution. Its introduction  may be necessary to
access to degrees of freedom which cannot be described in terms
of the $L^2$ modes (\ref{kgbase}). This important extension { to non-$L^2$  ``classical''   modes}
deserves a thorough examination and is incidentally mentioned here.

Eq.(\ref{general}) reduces to a Bogoliubov
transformation of the reference theory (\ref{kgbase}) only in the special case
(\ref{kgab}). These states are {\em pure states}.
The other states entering in the enlarged canonical formalism of Eq. (\ref{general})
are in general mixed states; for mixed states the representation
of the field algebra is not irreducible.

The discussion of Minkowski KMS states has pointed out that
simple but physical important examples of  two-point functions
have the following  structure:
\begin{equation}
\underline \W_{\a,\b}(x,x') = \sum [\a_{ij
}\a^*_{il} \, u_j(x) u^*_l(x') + \b_{ij}\b^*_{il} \, u^*_j(x) u_l(x')], \label{kgab2}
\end{equation}
with
\begin{equation}
\a_{ij}\a^*_{il} - \b^*_{ij}\b_{il} = \delta_{jl}.
\end{equation}
In particular, when $\a_{ij} = a_i \delta_{ij}$ and $\b_{ij} = b_i \delta_{ij}$ (no summmation intended) this condition is simply
$
|\a_i|^2 - |\b_i|^2 = 1.
$  and the two-point function
\begin{equation}
\underline \W_{\a}(x,x') = \sum \left[|\a_{i
}|^2\, u_i(x) u^*_i(x') + (|\a_{i}|^2-1) \, u^*_i(x) u_i(x')\right]. \label{kgab2bis}
\end{equation}
\newpage
\section{Quantum fields on the Rindler universe and the Unruh effect}

\subsection{Two dimensional Rindler Universe}

\subsubsection{Geometry}

The two-dimensional Rindler spacetime is nothing else that
{the right wedge
\begin{equation}\label{rwedge}
{\rm (R)} = \{ (t,\x) \in {\Bbb R}^2: \,|t|<\x\}
\end{equation}
 of a two-dimensional }Minkowski spacetime $M_2$ relative to  a chosen origin  $(t=0,x=0)$.
The Rindler coordinates are related to the inertial coordinates $(t,\x)$ {as follows
\begin{equation}
\left\{\begin{array}{lclcl} t &=& \r \sinh \e &=& \ex \sinh \e,\\
\x  &=& \r \cosh \e &=& \ex \cosh \e , \end{array}\right.
 \ \ \  \ \ \  (\xi,\eta) \in {\Bbb R}^2  \ \ \ \ \ \ \ \ \ \ \ \ {\rm(R)}            .
\label{csrw}
\end{equation}
The interval between }two events is given by
\begin{equation}
ds^2=\rho ^2\ {d\eta }^2 -{d\rho }^2 = e^{2 \xi }({d\eta}^2-d\xi^2), \\
\end{equation}\begin{equation}(x-x')^2 = 2 \rho  \rho' \cosh \left(\eta -\eta '\right)  -\rho ^2-{\rho '}^2 =2 \,e^{ \xi +\xi'} \cosh \left(\eta -\eta '\right)  -e^{2 \xi } -e^{2 \xi  '}.
\end{equation}

The  $\eta$ variable is interpreted as a global time coordinate for (R).
Physically  $\eta$ coincides with the proper time measured by an accelerated observer whose worldline {is the hyperbola  $t= \sinh \eta$, $ \ \x = \cosh \eta$ (i.e. $\rho = 1$ in Eq. (\ref{csrw})).
Translations }in the time coordinate $\eta \to \eta + a$ are isometries of the Rindler's manifold.
In inertial coordinates such time translations correspond to generic proper
Lorentz transformations and of course they leave the right wedge (R) invariant.

The coordinate system (\ref{csrw}) can be considered for complex values of {the coordinates  $\xi$ and $\eta$.
Complex events }  such that $\Im \rho = 0$ and  $\Im \eta > 0$ belong to the future tube
\begin{equation}\label{futuretube}
T^+ = \{z= x+iy\in { M}_2^{\rm{(c)}},\  y=\Im z \in V^+  \}
\end{equation}
of the complex Minkowski spacetime ${ M}_2^{\rm{(c)}}$ {where $V^+$ is the forward cone
\begin{equation}\label{futurecone}
V^+ = \{y\in { M}_2,\  y^2 = {y^0}^2-{y^1}^2 > 0,\ \ \  y^0>0  \in V^+  \}.
\end{equation}
Similarly, }events such that $\Im \rho = 0$ and  $\Im \eta < 0$ belong to the past tube $T^-$.
\subsection{Massless case: a review}
\label{maslessec}
The two-dimensional massless Klein-Gordon quantum field theory is special:
the conformal invariance of the two-dimensional massless field equation
gives rise to a very special kind of invariance of the modes
that have precisely the same form in both the Minkowski and the Rindler's coordinate systems.
 The massless case is specifically characterized by the infrared divergence of the
two-point functions that are obtained by canonical quantization.
Nonetheless,  the two-dimensional massless case encodes in a simple way all the features which ground
the common way of understanding the
Unruh and the Hawking effects.  The following short review, that closely follows the treatment in \cite{Birrell},
can help in putting our results in perspective.

The massless Klein-Gordon equation in standard coordinates
$$\Box \phi = \partial_t^2 \phi -\partial_x^2 \phi =0 $$
admits the following basis of positive frequency modes
\begin{equation}
 f_\k(t,\x)=\frac{1}{\sqrt{4\pi|\k|}}
e^{-i|\k| t+i\k\x}, \ \ \ \k \in {\bf R} \label{masslesswaves}
\end{equation}
that is complete and normalized in the Klein-Gordon sense (\ref{kgproduct})  w.r.t. (say)
the  Cauchy surface $\Sigma = \{(t,\x) \in {\Bbb R}^2: \ t=0\}$.

By using the above set of modes the (abstract) canonical quantum field can be
formally expanded in terms of creation and annihilation operators as in Eq. (\ref{field0}):
\begin{equation}
{\phi}(x)=\int_{-\infty}^{+\infty}\frac{d\k}{\sqrt{4\pi\left|\k\right|}}
\left[e^{-i\left|\k\right|t+i\k{\bf x}}
\,{a}{(\k)}+e^{i\left|\k\right|t-i\k{\bf x}}\,{a}^\dagger{(\k)}\right];
\label{pippo}
\end{equation}
this expression  implements the CCR's:
\begin{equation}
[{\phi}(x),\phi(x')]=\int_{-\infty}^{+\infty}\frac{d\k}{{2\pi\left|\k\right|}}\sin({-\left|\k\right|(t-t')+\k({\bf x-x'}})).
\label{ccruneq}
\end{equation}
Trying to realize these algebraic relations in a Hilbert space
{one faces a well-known difficulty: when computed in the vacuum vector $\left|0_{M}\right\rangle$ defined by the conditions}
\begin{equation}
\hat a(\k)\left|0_{M}\right\rangle = 0, \ \ \ \ \ \k\in {\bf R},
\end{equation}
the two-point expectation value of the  would be operator-valued distribution
\begin{equation}
{\hat \phi}(x)=\int_{-\infty}^{+\infty}\frac{d\k}{\sqrt{4\pi\left|\k\right|}}
\left[e^{-i\left|\k\right|t+i\k{\bf x}}
\,\hat{a}{(\k)}+e^{i\left|\k\right|t-i\k{\bf x}}\,\hat{a}^\dagger{(\k)}\right]
\label{fieldmasslessrep}
\end{equation}
{is  infrared-divergent:}
\begin{equation}
\langle 0_M|\hat \phi(x) \hat \phi(x')\left|0_{M}\right\rangle = \int dk f_\k(x) {f^*_\k(y)} =\frac{1}{4\pi}
 \int_{-\infty}^{+\infty}\frac{d\k}{|\k|}
e^{-i\left|\k\right|(t-s)+i\k{\bf (x-y)}} = \infty. \label{masslesstp1}
\end{equation}
Note however that the commutator (\ref{ccruneq}) is perfectly well-defined and free of infrared singularities.
There are different ways of regularizing the two-point function (\ref{masslesstp1}):
one can chose to keep positive-definiteness
at the price to obtain a non covariant quantization \cite{Klaiber}
or to maintain locality and covariance but the
quantization lives in an indefinite metric space  \cite{Klaiber,morchio};
in any case there does not exist any solution
of the functional equation (\ref{CR})
that is covariant {\em and} positive definite.

{The study of the same equation in Rindler coordinates}
\begin{eqnarray}  &&\Box \phi = \partial_\eta^2 \phi -\partial_\xi^2 \phi =0, \end{eqnarray}
gives rise to apparently
identical expressions; {the Rindler modes} and the abstract field $\phi$
have precisely the same functional dependence on the
coordinates and the same normalization coefficients as the previous expressions (\ref{masslesswaves}) and (\ref{masslesstp1})
{written in Minkowskian coordinates:
\begin{eqnarray} \
 && R_\k(\eta,\xi)= \frac{1}{\sqrt{4\pi |\k|}}
e^{-i|\k| \eta+i\k\xi}, \label{right}
\\
&&{\phi}(x)= \int_{-\infty}^{+\infty}\frac{d\k}{\sqrt{4\pi\left|\k\right|}}
\left[e^{-i\left|\k\right|\eta+i\k{\xi}}
\,{b}{(\k)}+e^{i\left|\k\right|\eta-i\k{\xi}}\,b^\dagger{(\k)}\right]. \label{fieldright}
\end{eqnarray}
Once more}, in the alternative vacuum vector defined by the conditions
\begin{equation}
\hat b(\k)\left|0_{R}\right\rangle = 0, \ \ \k\in {\bf R}
\end{equation}
the two-point expectation value of the field {also is  infrared-divergent}:
\begin{equation}
\langle 0_R|\hat \phi(x) \hat \phi(x')\left|0_{R}\right\rangle = \int d\k R_\k(x) {R^*_\k(x')} =\frac{1}{4\pi}
 \int_{-\infty}^{+\infty}\frac{d\k}{|\k|}
e^{-i\left|\k\right|(\eta-\eta')+i\k{(\xi-\xi')}} = \infty. \label{masslesstp}
\end{equation}
There is however an  obvious but important difference to be noted:
{the Rindler modes} (\ref{right}) are complete and normalized w.r.t. the Cauchy surface
$\Sigma' = \{(\eta,\rho) \in {\Bbb R}\times {\Bbb R}^+, \ \rho = 0\}\subset \Sigma$; however $\Sigma'$ is a
Cauchy surface only for {the Rindler manifold } (which is globally hyperbolic in itself)
but of course {\em not} for the Minkowski manifold.
As a consequence the family of modes (\ref{right}) is not a complete basis for the
Minkowski spacetime; on the other hand, for the Rindler spacetime usage the family
(\ref{masslesswaves}) is overcomplete and the modes
are not correctly normalized.

To make a comparison between the two set of modes one needs
to step out of the space under study, namely the right wedge (R), and to add another
part of Minkowski space, that is to add to our original space the other Rindler chart
$(\eta',\rho')$ which covers the left wedge (L):
\begin{equation}
\left\{\begin{array}{lclcl} t &=& -\r' \sinh \e' &=& -e^{\xi'} \sinh \e',\\
\x  &=& -\r' \cosh \e' &=& -e^{\xi'} \cosh \e',\end{array}\right. \ \ \ {\rm (L)},
\end{equation}
and then extend the ``right'' modes (\ref{right}) to (L) and complete the family  by adding the corresponding ``left'' modes:
\begin{equation}
R_\k(x) = \left\{\begin{array}{lcll}
\frac{e^{-i\left|\k\right|\eta+i\k{\xi}}}{\sqrt{4\pi\left|\k\right|}} & \makebox{in (R)},\\
  0 & \makebox{in (L)},
\end{array} \right. \,
\ \ \
L_\k(x)= \left\{\begin{array}{lcll}  0 & \makebox{in (R)}\, ,\\
\frac{e^{i\left|\k\right|\eta'+i\k{\xi'}}}{\sqrt{4\pi\left|\k\right|}} & \makebox{in (L)}.
\end{array}\right.  \label{extended}
\end{equation}
Correspondingly, for each $\k$ two more terms have to be added to the field expansion (\ref{fieldright})
to account for the creation and annihilation of the left modes.
{At this point the Bogoliubov transformation which relates the systems (\ref{masslesswaves}), in Minkowski coordinates,
and (\ref{extended}), in (double) Rindler coordinates, can be computed by a -not very enlightening- direct calculation.
A clever argument, which is due to Unruh \cite{unruh}, avoids that calculation;} it
is based on the observation that the following combinations
\begin{equation}\label{modesf}
f^{(1)}_\k = \frac{{e^{\frac{\pi |\k|}{2}}}R_\k + {e^{-\frac{\pi |\k|}{2}}}
\ {L}^*_{-\k} }{\sqrt{2\sinh{\pi|\k|}}},
\ \ f^{(2)}_\k = \frac{{e^{-\frac{\pi |\k|}{2}}}R^*_{-\k} +
{e^{\frac{\pi |\k| }{2}}}\ {L}_{\k} }{\sqrt{2\sinh{\pi|\k|}}},
\end{equation}
have the same analyticity properties of the waves
(\ref{masslesswaves}) in the tubes $T^+$ and $T^-$ (see \cite{Birrell} for details)
and therefore the Bogoliubov transformation between this basis
and the basis (\ref{masslesswaves}) {does not mix positive and negative Minkowski frequencies.}
Equivalently, by introducing the  annihilation and creation  operators $d_1(\k), \  d^\dagger_1(\k), \
 d_2(\k), \ d^\dagger_2(\k),$ relative to the modes (\ref{modesf})
{the Bogoliubov transformation which relates them to the operators $a(\k)$ and $a^\dagger(\k)$
does not mix the latter creation and annihilation operators.}
It follows that the Minkowski vacuum
is equivalently identified by the conditions
\begin{equation}\label{mv2}
\hat d_1(\k)|0_{M}\rangle = \hat d_2(\k)|0_{M}\rangle
\end{equation}
and the field expansion (\ref{fieldmasslessrep}) can be rewritten
\begin{equation}\label{fieldf}
 \hat{\phi}=\int_{-\infty}^{+\infty} \,\,
\left[f^{(1)}_\k \,\hat{d_1}(\k) + { \left(f^{(1)}_\k\right)^* }\,\hat{d_1}^\dagger(\k) +
 f^{(2)}_\k \,\hat{d_2}(\k) + \left(f^{(2)}_\k\right)^*\,\hat{d_2}^\dagger(\k)\right] d\k.
\end{equation}
By using Eq. (\ref{modesf}) we can now compute the Bogoliubov transformation between the two systems:
\begin{eqnarray}\label{bogorin}
b_L(\k) = \frac{{e^{\frac{\pi |\k|}{2}}}d_2(\k) + {e^{-\frac{\pi |\k|}{2}}}
\ {d}^\dagger_1{(-\k)} }{\sqrt{2\sinh{\pi|\k|}}}, \ \
b_R(\k) = \frac{{e^{-\frac{\pi |\k|}{2}}}d^\dagger_2({-\k}) +
{e^{\frac{\pi |\k|}{2}}}\ d_1(\k) }{\sqrt{2\sinh{\pi|\k|}}}.
\end{eqnarray}
According with the standard interpretation (see e.g. \cite{Birrell}){ the accelerated Rindler observers
see that the  Minkowski vacuum }contains $n_\k$ particles in the mode $\k$, where
\begin{equation}\label{unruheff}
n_\k = \langle 0_M|\hat b_R^\dagger(\k) \hat b_R(\k)|0_{M} \rangle = \frac{e^{-\pi |\k|}}{2\sinh{\pi|\k|}} =\frac1{e^{\pi |\k|}-1}
\end{equation}
i.e. the accelerated observer perceives a thermal radiation with  Planck spectrum.
{ This fact is  general and model independent:
restricting a Wightman quantum field theory to a wedge always gives rise to a KMS thermal equilibrium state} \cite{sewell,bisognano}.

\subsection{Massive case}
The discussion of the Unruh effect
for the massless Klein-Gordon field in two-dimensions
is plagued by the infrared divergence
of the two-point functions;  this feature invalidates
the formal treatment given in the previous section.

A discussion of the massive case
based on the ingredients, namely standard textbook
canonical quantization supplemented by the theory of Bogoliubov transformations,
would be much more complicated than in the massless case because
conformal invariance is missing in the massive case.

{We believe that there is an even more serious drawback in the method used above:
the necessity to extend} the Rindler model to the left wedge
is neither economical from a theoretical viewpoint
nor can be applied in more general situations where the
required extension might not be so obvious: just consider that it took fifty years to discover the
Kruskal extension of the Schwarschild metric.

In this section
we give a full description of the massive two-dimensional
case by applying the extended canonical formalism \cite{Moschella:2008ik}.
{ The method that we are going to present in this paper does not require to know
any "extension" at all: it has the great advantage of working solely
within the right wedge which models the Rindler universe.

Consider  a $d$-dimensional Minkowski spacetime $M_d$.}
The Rindler coordinates that cover the right wedge are
\begin{equation}
\left\{\begin{array}{lclcl} x^0 &=& \r \ \sinh \e &=& \ex \sinh \e\\
x^i &=& \x^i\\
x^{d-1} &=& \r\ \cosh \e &=& \ex  \cosh \e\end{array}\right.
\end{equation}
The interval is then written
\begin{equation}
ds^2=\rho ^2\ {d\eta }^2- {d\rho }^2  -d {\x} ^2 = e^{2 \xi }({d\eta}^2-d\xi^2)-d {\x} ^2
\end{equation}
The  Klein-Gordon equation appears as follows:
\begin{equation}
e^{-2\xi}(\partial^2_\eta \phi-\partial^2_\xi \phi)-\triangle\phi + m^2 \phi= e^{-2\xi}(\partial^2_\eta u-\partial^2_\xi u)- (\k^2+m^2 ) \phi= 0.
\end{equation}
In the second step we have inserted a factorized solution of the form
\begin{equation}
\phi(x) = u_\k(\eta,\xi) e^{i\k \cdot\x} .
\end{equation}
Clearly the general $d$-dimensional case will follow from the study two-dimensional
massive Klein-Gordon equation
\begin{equation}
\partial^2_\eta \phi-\partial^2_\xi \phi + m^2 e^{2\xi}\phi = 0\, .
\end{equation}
simply by the shift $m\to \sqrt{m^2+k^2}$ wherever the mass parameter appears
and by multiplication of the ``transverse'' waves $e^{i\k\x}$.
We therefore focus on the two-dimensional case and  look  for solutions  of the form
$
u(\eta,\xi) \simeq e^{-i\omega \eta}F_{\omega}(\xi), \ \omega\geq 0$,
which are positive frequency  w.r.t. the conformal time; the function $F$ has to satisfy  the modified Bessel equation
\begin{equation}
-\partial^2_\xi F + m^2 e^{2\xi}F= \omega^2 F .
\end{equation}
The solution that behaves well at infinity is the modified Hankel-Macdonald function $K_{i\omega}(me^\xi)$.
A spacelike surface that may be used to compute the normalization is for instance the
half-line $\eta= \eta_0$, $\xi \in {\bf  R}$.  The result does not depend on the choice of one
particular half line because they all share the same origin. In doing this we
are applying the standard canonical formalism in the Rindler
wedge. To get the normalization we
note that for  $\varepsilon>0$
\begin{eqnarray}
&& \int t^{-1+\varepsilon} K_{i\omega_1}(t)K_{i\omega_2}(t)dt = \frac
{ \left| \Gamma \left(\frac{1}{2} (\varepsilon +i\omega_1 + i\omega_2)\right)\right|^2
 \left| \Gamma \left(\frac{1}{2} (\varepsilon +i\omega_1 - i\omega_2)\right)\right|^2 }{2^{3-\varepsilon}\Gamma (\varepsilon)} \cr && \cr
&& \simeq \frac{\pi} {(\omega_1+ \omega_2)\sinh \frac 12 \pi (\omega_1+ \omega_2)}\left(\frac{\varepsilon}
{\varepsilon^2 + (\omega_1-\omega_2)^2} \right)
 \mathop{\longrightarrow}_{\varepsilon\to 0} \ \
\frac{\pi^2}{2 \omega_1 \sinh \pi \omega_1}\delta (\omega_1-\omega_2). \cr &&
\end{eqnarray}
Therefore, a  convenient orthonormal system solving (\ref{kgbase}) for the massive Klein-Gordon
equation in the Rindler universe can  be   written  as follows:
\begin{eqnarray}
\left\{\begin{array}{lll}u_\omega(\eta,\xi) &=&\frac{\sqrt{\sinh \pi \omega}}{\pi}
e^{-i\omega \eta}K_{i\omega}(me^\xi)\;\;\cr
u^*_\omega(\eta,\xi) &=&\frac{\sqrt{\sinh \pi \omega}}{\pi}
e^{+i\omega \eta}K_{i\omega}(me^\xi)\;\; \ \
\end{array} \right. \omega>0.\label{rindlermodes}
\end{eqnarray}
The modes $u_\omega(\eta,\xi)$  are positive frequency w.r.t. the Rindler's time coordinate $\eta$.
The algebraic massive canonical quantum Klein-Gordon field can thus { be written in  Rindler coordinates as follows:}
\begin{eqnarray}
\phi(x)
&=& \frac 1\pi  \int_0^\infty
\left[ e^{-i\omega \e} \, a(\omega) + e^{i\omega\eta}\,
\,a^\dagger(\omega)\right] K_{i\omega}(m \ex)\sqrt{\sinh \pi \omega}\ d\omega
\label{field}.
\end{eqnarray}
and the so-called  Fulling vacuum \cite{Fulling:1972md,Fulling:1977zs} is identified by the condition
\begin{equation}\label{fullingvac}
\hat a(\omega) |0_R \rangle = 0, \ \ \ \ \ \omega \geq 0.
\end{equation}
The  two-point function for the massive Klein-Gordon field represented
in  the Fulling vacuum  is given by
\begin{equation}
W_R(x,x')= \langle 0_R |\, \hat \phi(x) \hat \phi(x') |0_R \rangle =
\frac 1{\pi^2}\int_0^\infty e^{-i\omega (\e-\e') }  \,
 K_{i\omega}(m \ex) K_{i\omega}(m e^{\xi'}) {\sinh \pi \omega}\ d\omega. \label{rindler1}
\end{equation}

{We  now show how our formalism allows for a direct}
construction of the Wightman vacuum  (\ref{tp0})
characterized by Poincar\'e invariance and the standard spectral condition
working solely within the right Rindler wedge and using only the
``right'' modes (\ref{rindlermodes}) thus avoiding the need
of extending the system to the left wedge.

{ In the first step, we insert the modes (\ref{rindlermodes}) in Eq. (\ref{general}), with $S(x,y)=0$, to get
\begin{eqnarray}
W(x,x') &=&   \int_0^\infty d\omega \,  \int_0^\infty d\omega'     \,
 \left[ \delta_{\omega,\omega'} + B_{\omega,\omega'} \right] u_\omega(\eta,\xi) u^*_{\omega'}(\eta',\xi') \cr
 &+&   \int_0^\infty d\omega \,  \int_0^\infty d\omega'     \, B_{\omega,\omega'}  \, u^*_\omega(\eta,\xi) u_{\omega'}(\eta',\xi') \cr
 \label{rindlergen}
&+&   \int_0^\infty d\omega \,  \int_0^\infty d\omega'     \, \Re \,\, \{ C_{\omega,\omega'} \, \left[u_\omega(\eta,\xi) u_{\omega'}(\eta',\xi') + u_\omega(\eta',\xi') u_{\omega'}(\eta,\xi) \right] \, \}. \cr &&
\end{eqnarray}
This provides for a large -much larger than in the standard approach- family of mathematically}
admissible two-point functions (and therefore representations) for the massive Rindler Klein-Gordon field,
all of them sharing the same commutator $C(x,x')$ and, a fortiori, the canonical equal time commutation relations.

In the second step, we select those theories
in which the wedge-preserving Lorentz boosts $\eta \rightarrow \eta + a$ are unbroken symmetries;
{the latter condition requires to match frequencies, which in turn imposes  the following restrictions on (\ref{rindlergen}):
\begin{equation}
B_{\omega,\omega'}  = \left[\alpha(\omega)-1\right] \, \delta_{\omega,\omega'}, \ \ \ \ C_{\omega,\omega'} = 0.
\end{equation}
At this point, }we have constructed a family of representations parameterized
by the choice of an arbitrary function $\alpha(\omega)\geq 1$;
they are associated to the following two-point functions (see Eq. (\ref{kgab2bis})):
\begin{eqnarray}
W_{(\alpha)}(x,y) =
\frac 1{\pi^2}\int_0^\infty [\alpha(\omega) e^{-i\omega (\e-\e')} + (\alpha(\omega)-1) e^{+i\omega (\e-\e')}]  \times   \,
 \cr \times K_{i\omega}(m \ex) K_{i\omega}(m e^{\xi'}) {\sinh \pi \omega}\ d\omega  \label{rindler1a}
\end{eqnarray}
The function $\alpha(\omega)$ is assumed to be smooth (with the possible exception of $\omega = 0$)
and to have a limit at infinity:
\begin{equation}\label{klkllk}
\lim_{ \omega \to \infty}\alpha(\omega) = \ggamma\geq 1.
\end{equation}

Any choice of $\alpha(\omega)$  satisfying the above properties gives rise to
a plausible Lorentz invariant (i.e. invariant under Rindler'time translations)
canonical quantization of the massive Klein-Gordon field on the Rindler manifold.
Can we construct theories that are Poincar\'e invariant?
Of course the Rindler manifold is not globally Poincar\'e invariant.
But there is the possibility to act with Poincar\'e transformations locally.

\subsubsection{Enforcing Poincar\'e invariance}
{There is no loss of generality in extending  the weight function  to negative values of
the Rindler energy variable $\omega$ by defining:
\begin{equation}
\alpha(-\omega)  = 1- \alpha(\omega).
\end{equation}
It is also useful to write
\begin{equation}
\alpha(\omega) = \ggamma f(\omega) + (1-\ggamma) f(-\omega)
\end{equation}
where  $f(\omega)$  is such that
\begin{equation}\label{gag}
f(\omega) + f(-\omega)= 1, \  \ \   \lim_{\omega\to \infty}  f(\omega) = 1 .
\end{equation}
The two-point function  (\ref{rindler1a}) can then be rewritten as follows:
\begin{eqnarray}
W_{(\alpha)}(x,x')  =
\frac 1 {\pi^2}\int_{-\infty}^\infty {\sinh \pi \omega} \ [ \ggamma  f(\omega) + (1-\ggamma) f(-\omega) ] \ e^{-i\omega (\e-\e')} \times \cr   \,
\times  K_{i\omega}(m \rho) K_{i\omega}(m \rho') \ d\omega. \label{rindler1ter}
\end{eqnarray}
To handle the integration in (\ref{rindler1ter}) suitable integral representations of the modified Hankel
function are needed. Two useful representations can be derived from the following general expression:
\begin{eqnarray}
K_{i\omega}(z)
= K_{-i\omega}(z)=\frac12\,  \int_{0}^\infty  \ e^{-\frac 12 z(t+t^{-1})} t^{-i\omega} \frac{dt}{t}, \ \ \  \Re z >0 \label{expression} . \label{hankel}
\end{eqnarray}
By contour distortion the integration may be done over the path $t \to te^{i\theta}$
with $-\frac{\pi}2 <\theta< \frac \pi 2$; the integral (\ref{hankel}) stays convergent
provided that $|\Im z \sin \theta |<\Re z \cos \theta$.

The two relevant integral representations of the Hankel-Macdonald
arise in the two limiting cases
$\theta = \pm \frac \pi 2$ where the integrals are only marginally convergent;
the representations are valid only for real and positive values of the argument.
Changing to the variable $t = \exp \lambda$ we get:
\begin{eqnarray}
K_{i\omega}(m\rho)
=\frac12\,e^{-\frac 12 \pi \omega}  \int_{-\infty}^\infty   e^{ i m\rho \sinh \lambda}
e^{-i\omega \lambda} {d\lambda}\ , \ \ \rho>0,\label{expressionx1}\\
 K_{i\omega}(m\rho)
 = \frac12\,e^{\frac 12 \pi \omega}  \int_{-\infty}^\infty   e^{- i m\rho \sinh \lambda}
e^{-i\omega \lambda} {d\lambda}\ ,\ \  \rho>0. \label{expressionx2}
\end{eqnarray}
By inserting the above representations in Eq. (\ref{rindler1ter})
we obtain the final expression for the two-point function:
\begin{eqnarray}
W_{(\alpha)}(x,x') =
\frac \ggamma {4 \pi^2}\int_{R^3} F(\omega) \ e^{-i\omega (\e-\e'+ \lambda - \lambda')}    \,
e^{im \rho \sinh \lambda- im \rho' \sinh \lambda'} \ d\omega d\lambda d\lambda'+ \cr \\ \label{rindler131}  +
\frac {\ggamma-1} {4\pi^2}\int_{R^3}  F(-\omega) \ e^{-i\omega (\e-\e'+\lambda - \lambda')}    \,
e^{-(im \rho \sinh \lambda- im \rho' \sinh \lambda')}  \ d\omega d\lambda d\lambda'
\nonumber
\end{eqnarray}
where we have defined
\begin{equation}
F(\omega) =  e^{-\pi\omega} \, {\sinh \pi \omega}\, f(\omega).
\end{equation}
The two-point function (\ref{rindler131}) is manifestly Lorentz invariant,
as it depends  only on the difference of the temporal coordinates $\eta-\eta'$.
Let us examine the variation of Eq. (\ref{rindler131}) under infinitesimal spacetime translations:
\begin{eqnarray}
&& \left\{\begin{array}{lclrlrr} \r\ \delta \e  &=&   \ \cosh \e \ \delta  x^0 - \sinh\e \ \delta x^1  \\
 \delta \r  &=&  - \ \sinh \e \ \delta  x^0 +\cosh \e \ \delta x^1    \end{array}\right. ,\\ && \cr
 && \left\{\begin{array}{lclrlrr}  \r'\ \delta \e'  &= &   \ \cosh \e' \ \delta  x^0 - \sinh\e' \ \delta x^1\\
 \delta \r'  &=&  - \ \sinh \e' \ \delta  x^0 +\cosh \e' \ \delta x^1 \end{array}\right. \, .
\end{eqnarray}
Consider in particular the infinitesimal space translations
\begin{equation}
\delta x^0=0, \ \
\delta x^1 =\epsilon>0
\end{equation}
that map the right wedge into itself.
Requirement of invariance of (\ref{rindler131}) under such transformations imposes that the following variation
must vanish for every choice $\rho,\eta$ and $\rho',\eta'$:
\begin{eqnarray}
\delta W_R =
\frac {im\epsilon\ggamma}{4 \pi^2} \int_{{\bf R}^2} \tilde F(\lambda - \lambda')   \,
 \,\left[\sinh \lambda-\sinh \lambda'\right]  e^{im \rho \sinh (\lambda-\eta)-
 im \rho' \sinh (\lambda'-\eta')} \ d\lambda d\lambda'
 \cr
\\Ê\label{rindler13}
+
\frac {im\epsilon(1-\ggamma)} {4\pi^2} \int_{{\bf R}^2} \tilde F(\lambda' - \lambda)    \,
\left[{\sinh \lambda-\sinh \lambda'}\right] e^{-im \rho \sinh
(\lambda-\e)+ im \rho' \sinh (\lambda'-\e')}  d\lambda d\lambda' ,
\cr
\cr
\nonumber
\end{eqnarray}
\begin{equation}\label{variation}
\delta W_R =0 \ \ \ \ \ \ \forall \rho,\,\eta,\, \rho',\,\eta'
\end{equation}
where $\tilde F(\lambda)$
is the Fourier transform of $F(\omega)$ that exists in some distribution
space of type ${\cal S}$ \cite{gelfand}  }.
The vanishing of the variation $\delta W_R$
provides an integral equations for the function $F(\omega)$ that depends parametrically on
$\rho,\eta$ and $\rho',\eta'$ and must hold for every choice of $\rho,\eta$ and $\rho',\eta'$.
Any solution of (\ref{variation}) provides a Poincar\'e invariant quantization expressed in Rindler's coordinates.

\subsubsection{Poincar\'e invariant states and the Unruh effect}
The solution of Eq. (\ref{variation}) valid for any  real $\ggamma\geq 1$ can be indeed found immediately
in the Schwartz space of tempered distributions ${\cal S}'(\Bbb R)$:
\begin{equation}
\tilde F(\lambda-\lambda') = c \delta(\lambda-\lambda'). \label{delta}
\end{equation}
The value of the constant $c$ can be determined by applying the last requirement
in Eq. (\ref{gag}) to the Fourier antitransform of (\ref{delta}):
\begin{equation}\label{kokonuda}
F(\omega)= \frac{c}{2\pi} = \frac{1}{2}, \ \ \ \ f(\omega) = \frac{e^{\pi \omega}}{2\sinh \pi \omega};
\end{equation}
the above value of the constant $c$ guarantees that the CCR's hold with the correct coefficient.
Finally, we obtain a family of two-points functions labeled by a real constant $\ggamma\geq 1$ (see Eq. (\ref{klkllk})
which fulfill  the variational condition  (\ref{variation})
\begin{equation}\label{huizinga}
\alpha(\omega) = \frac{\ggamma e^{\pi \omega} + (\ggamma-1) e^{-\pi \omega}}{2\sinh \pi \omega}
\end{equation}
For any chosen value $\ggamma\geq 1$ the two-point function provides a canonical  quantization
of the Klein-Gordon field in the Rindler's universe which is
furthermore Poincar\'e invariant and can therefore
be naturally extended to the whole two-dimensional Minkowski spacetime.
It may be more suggestive to introduce  another parameter $\zeta$ as follows
\begin{equation}\label{huizinga2}
\ggamma = \frac{e^{\pi\zeta}}{2\sinh \pi \zeta}= \frac{1}{1-e^{-2\pi\zeta}}
\end{equation}
so that
\begin{equation}\label{huizinga3}
\alpha(\omega) = \frac{\cosh(\pi \omega+\zeta )}{\cosh(\pi \omega+\zeta )- \cosh(\pi \omega-\zeta )}
\end{equation}
\begin{eqnarray}
W_{\frac{1}{\zeta}}(x,x')  =
\frac 1 {2\pi^2\sinh\beta}\int_{-\infty}^\infty
 \ e^{-i\omega (\e-\e')} \ {\cosh{\pi (\omega+\zeta)}} \ K_{i\omega}(m \rho) K_{i\omega}(m \rho') \ d\omega. \label{rindler1tergo}
\end{eqnarray}
In particular for $a=1$ i.e. $1/\zeta=0$ we have
\begin{eqnarray}
W_{0}(x,x') =
\frac 1{ \pi^2}\int_0^\infty
\ K_{i\omega}(m \rho) K_{i\omega}(m {\rho'})\cosh{\omega (\pi - i \e-i \e' )} d\omega = \label{rindlerpos} \\ =
\frac 1{2 \pi}  K_{0}\left(m \sqrt{\rho^2+{\rho'}^2- 2\rho\rho'\cosh (\eta-\eta')}\right) = \frac 1{2 \pi}  K_{0}\left(m \sqrt{-(x-x')^2} \right)
.
\end{eqnarray}
We have recovered this way the standard Poincar\'e-invariant
two-point function of a massive Klein-Gordon field with
positive energy spectrum.  The Unruh interpretation  follows from
by now classic arguments based on the Bisognano-Wichmann theorem \cite{bisognano,sewell}
or else may be explicitly reconstructed following the description given in the massless case.
The other solution corresponding to finite values of $\zeta$ contain negative (Minkowski)
energies and are nothing else but the two-point functions (\ref{nondiag2}) expanded in the  Rindler modes (\ref{rindlermodes}).

There are other theories having a special status in the family (\ref{rindler1a}),
which we recall is already a subset of the general family (\ref{general}).
The most noticeable example is the one-parameter family of states identified by the choice
\begin{equation}
\alpha(\omega) = (1-e^{-\beta \omega})^{-1} ,    \ \ \ \ \beta > 0.
\end{equation}
Let us  write the corresponding two-point function explicitly:
\begin{equation}
\W_{\beta}(x,y) =
\frac 1{ \pi^2}\int_0^\infty \left[ \frac{e^{-i\omega (\e-\e') }}{1-e^{-\beta\omega}} +    \frac{e^{i\omega (\e-\e')}}{ e^{\beta \omega}-1}\right]
K_{i\omega}(m \ex) K_{i\omega}(m e^{\xi'}) {\sinh \pi \omega}\ d\omega.  \label{rindler1c}
\end{equation}
Since $K_{i\omega} = K_{-i\omega}$  and since $|K_{i\omega}(\rho) K_{i\omega}(\rho') {\sinh \pi \omega}|$ is bounded at infinity in the $\omega$ variable,
one can immediately  check that $\W_{\beta}(x,y)$ verifies the KMS analyticity and periodicity  properties in imaginary time \cite{haag,Bros:1992ey} at inverse temperature
$\beta$.  These states are precisely  the KMS states in Rindler space, which have been introduced and   characterized in  \cite{kay,kay2}.
The special value $\beta = 2\pi$ has also been identified \cite{kay} with the restriction to the wedge of the Wightman vacuum on the basis
of the Bisognano-Wichmann and Reeh-Schlieder theorems. Our proof follows just by enforcing the requirement that the wedge preserving
translation be an exact symmetry.



\section{Summary and concluding remarks\label{sec:concl}}

There is much more flexibility in canonical quantization than
is usually believed.
The simple but very general modification of the
standard formalism that we have described
opens a vast class of new possibilities for constructing
canonical fields by means of the Fock construction.

When a ``complete and orthonormal'' set of modes is available
our extension of the canonical formalism
amounts to considering the most general quadratic combination
of the modes that is compatible with a given
commutator, as prescribed in the fundamental functional equation (\ref{CR}).
For Klein-Gordon fields our generalization is  given in Eq. (\ref{general}).
Similar formulae can be constructed for spinor or tensor fields.

Our scheme  produces interesting results already for
quantum fields in the flat Minkowski case.
We have for instance described a simple and original
derivation of the KMS thermal equilibrium states, { but also unraveled a
wealth of other similar }unexplored possibilities as well.

Another situation where our extended canonical formalism may be useful
is commonly encountered in concrete examples when
the coordinates systems {that one wants (or is compelled) to use}  cover only a
portion of the manifold under study.
In such cases the spacelike hypersurfaces used for canonical quantization
are not in general Cauchy surfaces for the extended manifold;
the patch covered by the coordinate system may or may not
be a globally hyperbolic manifold in itself.

For example, this situation is encountered
in black-hole spacetimes or in the Rindler coordinate system
of a wedge of a Minkowski spacetime \cite{hawking,unruh,sewell,bisognano,kay,kay2,fulling,kayw}.
The same phenomenon may also happen in cosmological backgrounds.
Depending on the behavior of the scale factor $a(t)$, the surface  $t=${\em constant} may fail to
be a Cauchy surface for the maximally extended manifold
(while it is a Cauchy surface for the patch covered by the Friedmann-Lema\^{\i}tre-Robertson-Walker coordinates).

In this circumstances textbooks suggest  the
use of the standard canonical formalism \cite{Birrell}, but many possible
quantizations are lost in this very initial step.

{We  have given some simple examples how our generalization of the canonical formalism can be used;
 among other things we have given a fresh and comprehensive discussion of QFT in the Rindler wedge  revisiting  in particular the Unruh effect.}
The important
characteristic of our approach is its ability to reconstruct
the standard Poincar\'e invariant vacuum working solely within
the Rindler wedge.
Only local considerations are required to get the
globally invariant vacuum,
with no need to perform analytic continuation to
(or from) whatever ``external''  spacetime region which
may be added.

Another example which shares some of the features of the Rindler model is
the open de Sitter universe that has been treated in \cite{Moschella:2008ik} (see also \cite{MS,MS1}).
Also in this case the coordinate system gives access only to a
portion of the manifold; while canonical quantization in these
coordinates (plus Bogoliubov transformations)
produces theories that are not de Sitter invariant, application
of our procedure gives rise to the preferred de Sitter
invariant theory \cite{gibbons,BM}.
When the mass is lower than a critical value there are also
non-standard non-square integrable modes which come into the
play (see Eq. \ref{general}). These modes were known to exist, but their physical
relevance was quite uncertain. We have seen here that such contributions
are not at variance with the principles of quantum mechanics
since they do not contribute to the commutator.
{This novelty will be discussed in a forthcoming work  \cite{usexp}.}



\bibliographystyle{aipproc}   


\end{document}